\documentclass{raa}
\usepackage{graphicx,times}             %for PS/EPS graphics inclusion, new
\usepackage{txfonts}
\usepackage{color}
\usepackage{url}

\newcommand {\Msun}        {\mbox{M$_{\odot}$}}
\newcommand {\Rhm}         {\mbox{$R_\mathrm{hm}$}}

\begin{document}

   \title{GalevNB: a conversion from N-BODY simulations to
   observations
%\footnotetext{$*$ Supported by the National Natural Science Foundation of China.}
}
%   \subtitle{I. Place Your Subtitle Here}

   \volnopage{Vol.0 (200x) No.0, 000--000}      %%preserved for Editor. DOn't remove!
   \setcounter{page}{1}          %%starting page, preserved for Editor. DOn't remove!

   \author{Xiaoying Pang
      \inst{1,2,5}
   \and Christoph Olczak
      \inst{1,3}
   \and Difeng Guo
     \inst{1,3}
   \and Rainer Spurzem
      \inst{1,4,3,6}
   \and Ralf Kotulla
      \inst{7}
   }

   \institute{National Astronomical Observatories and Key Laboratory of Computational Astrophysics, Chinese Academy of
Sciences, 20A Datun Road, Chaoyang District, 100012 Beijing, P.R. China; {\it xypang@bao.ac.cn}\\
        \and
             Shanghai Institute of Technology, 100 Haiquan Road,
Fengxian district, Shanghai 201418, P.R. China\\
        \and
             Astronomisches Rechen-Institut, Zentrum f\"ur
Astronomie der
          Universit\"at Heidelberg, M\"onchhofstr.\ 12--14, 69120
              Heidelberg, Germany\\
         \and
         Kavli Institute for Astronomy \& Astrophysics, Peking University,
                 5 Yi He Yuan Road, Hai Dian District, Beijing 100871, P.R. China
         \and
Key Lab for Astrophysics, Shanghai Normal University, 100 Guilin
Road, Shanghai 200234, P.R. China
        \and
        Key Laboratory of Frontiers in Theoretical Physics, Institute of
Theoretical Physics, Chinese Academy of Sciences, Beijing, 100190,
P.R. China
       \and
Department of Astronomy, University of Wisconsin - Madison, 475 N Charter St, Madison WI 53706, USA
   }

   \date{Received~~2014 month day; accepted~~2014~~month day}

\abstract{We present {\tt GalevNB} ({\tt Galev} for $N$-body
simulations), an utility that converts fundamental stellar
properties of $N$-body simulations into observational properties
using the {\tt GALEV} (GAlaxy EVolutionary synthesis models)
package, and thus allowing direct comparisons between observations
and $N$-body simulations. It works by converting fundamental stellar
properties, such as stellar mass, temperature, luminosity and
metallicity into observational magnitudes for a variety of filters
of mainstream instruments/telescopes, such as HST, ESO, SDSS, 2MASS,
etc.), and into spectra that spans from far-UV (90 $\rm \AA$) to
near-IR (160 $\rm \mu$m). As an application, we use {\tt GalevNB} to
investigate the secular evolution of spectral energy distribution
(SED) and color-magnitude diagram (CMD) of a simulated star cluster
over a few hundred million years. With the results given by {\tt
GalevNB} we discover an UV-excess in the SED of the
cluster over the whole simulation time. We also identify four
candidates that contribute to the FUV peak, core helium burning
stars, thermal pulsing asymptotic giant branch (TPAGB) stars, white
dwarfs and naked helium stars.
 }
    \authorrunning{X. Y. Pang \& C. Olczak \& D. F. Guo \& R. Spurzem }            %author_head in even pages
   \titlerunning{GalevNB: a conversion from \texttt{NBODY} simulations to observations }  % title_head in odd pages

   \maketitle
%% The author head (on even pages) and the title head (on odd pages) will be
%% automatically extracted from \author{} and \title{}. Whenever the title is too long,
%% you will be asked to supply a shorter one by inserting either \authorrunning{} or
%% \titlerunning{} before \maketitle. Anyway, you can specify your own heads.
%%
%%
%% Note: In the following text body of your manuscript, please note several differences from
%%       other major journals:
%% (1) \subsection{Please Capitalize the First Letter of Each Notional Word in Subsection Title}
%% (2) Please Capitalize the First Letter of Each Notional Word in all tables' captions

%
%________________________________________________ sections below
%
\section{Introduction}           %% first-level sections will be auto-capitalized

Models of dense stellar systems, such as globular or young dense stellar clusters,
or nuclear star clusters with or without massive central black hole, require
direct $N$-body simulations, which at least resolve and follow all
stellar orbits in a star cluster as precisely as possible given the
usable hardware and computing time. But due to the strong
dependence on initial conditions of the gravitational $N$-body problem, 
which originates from close encounters, the
system is physically and numerically unstable (Miller 1966).
Therefore results of direct high-accuracy $N$-body simulations
should always be carefully analyzed, as they represent individual
realizations of physically possible solutions. In another word,
small variations of initial conditions may lead to different results
with regard to individual objects or events. Nevertheless, the
coarse-grained evolution (densities, distribution functions) is
quite well following expectations from statistical mechanics (see
e.g. Giersz \& Spurzem 1994, Spurzem \& Aarseth 1996). And the simulations
are an excellent tool to examine and study physical processes in
star clusters.

The sequence of high-accuracy direct $N$-body codes developed
by Sverre Aarseth (Aarseth 1999; Aarseth 2003) is among the most
widely used if not {\em the} most widely used code.
Its most notable features are the Hermite 4$^{\rm th}$ order integration scheme,
the hierarchically blocked individual time steps (Makino \& Aarseth
1992), the Kustaanheimo-Stiefel (KS) regularization for strong
interactions (Kustaanheimo \& Stiefel 1965) and its generalizations
for few-body subsystems (chain regularization, Mikkola \& Aarseth
1993), and the Ahmad-Cohen neighbor scheme (Ahmad \& Cohen 1973).
\texttt{NBODY6++} is the current massively parallel version of \texttt{NBODY6}
(Spurzem 1999, Spurzem et al. 2008), and 
\texttt{NBODY6++GPU} adds to it the use of accelerating many-core hardware
(GPU) on every node (Wang et al. 2015). In the cited paper it becomes
clear that direct $N$-body simulations with a million particles are
feasible now.
The challenges of the resulting
massive simulation data and their possible approaches of management
are addressed in \cite{cai14}.

However, the output parameters of N-BODY simulations are
mostly theoretical values. To make a direct comparison between our
simulation data and observations, we combine \texttt{GALEV} (GALaxy
EVolutionary synthesis models; Kotulla et al. 2009),  a flexible
algorithm to combine astrophysical colors in many filters and
spectra of stars (Lejeune, Cuisinier \& Buser 1997, 1998) or sets of
stars, with \texttt{NBODY6++GPU} simulations. In this paper, we present
this new code: \texttt{GalevNB} ({\tt Galev} for $N$-body
simulation) and its application in \texttt{NBODY6++GPU} simulation
data. Adapting subroutines from \texttt{GALEV}, \texttt{GalevNB} can
produce spectra spans the range from far-UV (FUV) at 90\,$\rm \AA$
to far IR at 160\,$\mu$m, with a spectral resolution of 20\,$\rm
\AA$ in the UV-optical and 50-100\,$\rm \AA$ in the near IR range.
Given a list of requested filters in HST, ESO, SDSS, 2MASS etc.,
\texttt{GalevNB} convolves the spectra with the filter response
functions and applies the chosen zero-points (Vegamag, ABmag, and
STmag) to yield absolute magnitudes. \texttt{GalevNB} bridges
theoretical parameters and their observed values, thus allows us to
understand the color and spectra evolution of star clusters, and to
determine the initial conditions and parameters of star cluster
simulations with a direct comparison to observations.

Though most Galactic and many Local Group clusters are resolved, observing individual stars in extragalactic star clusters is extremely
challenging due to severe crowding. Therefore, integrated photometry and spectroscopy
have been used to identify extragalactic star clusters and obtain
parameters (Bica et al. 1996a, Sarajedini et al. 2007,
Peng et al. 2008, 2009, Johnson et al. 2012), and even to investigate multiple stellar populations
 (Peacock et al. 2013). Using the imaging obtained with the High Resolution
Channel of the Advanced Camera for Surveys on board HST, 
Larsen et al. (2011) manage to construct color magnitude
diagrams (CMDs) for several extragalactic
star clusters within 5\,Mpc. However, for the very distant star clusters, such as the
ones in Antennae galaxies
($\rm \sim22\,Mpc$, Whitmore et al. 2007, Bastian et al. 2009), or in Virgo cluster ($\rm \sim16.5\,Mpc$, Peng et al. 2008, Peng et al. 2009),
studies of individual stars are still not possible. 
Integrated colors can be used to derive the age of star clusters 
generally (Elson \& Fall 1985). However, metal variation, stochastic effect (Girardi
et al. 1995), and dynamical evolution (Fleck et al. 2006) will change integrated colors. 
Therefore, integrated photometry may mot be enough to decouple the
stellar and dynamical effects of distant star clusters. 
With \texttt{GalevNB} producing observational magnitudes and spectra for
$N$-BODY simulations, it allows us to investigate stellar evolution and dynamics (via
colors and spectra) in distant star clusters at the same time, 
and even make prediction
for observations.   

% used simplified stellar dynamical models coupled
%with the same stellar evolution models as they are used here to show
%how dynamical evolution changes colours of star cluster in integrated
%light. They early used spherically symmetric gaseous models of star
%clusters, no N-BODY simulation. 
%Early in 1995, Girardi et al. (1995)

% 3) prediction of future evolution of star
%clusters; 4) age determination of observed clusters.

In this paper, we carry out star cluster simulations with
\texttt{NBODY6++GPU} codes, and use \texttt{GalevNB} to produce
observational data for the simulated cluster. The computational
methods of \texttt{NBODY6++GPU} simulations are summarized in Section
\ref{sec:numerical_method}. An introduction to the structure and
execution procedure of \texttt{GalevNB} is presented in Section
\ref{sec:galevnb_sim}. The observational features of the simulated
cluster, CMDs and spectral energy
distributions (SEDs) are outlined in Section \ref{sec:obs}. Finally,
we present our discussion and summary in Section \ref{sec:dis}.

\section{Computational Method and Initial Conditions}

\label{sec:numerical_method}

The model clusters in this work are evolved using \texttt{NBODY6++GPU}
which is the \texttt{MPI} parallel version based on the
state-of-the-art direct $N$-body integrator \texttt{NBODY6GPU}
(Aarseth 2003, Nitadori \& Aarseth 2012).
 Gravitational forces are computed using a fourth-order Hermite integration scheme with block time
steps (and without softening). 
%The additional spatial decomposition
%via the \emph{Ahmad-Cohen neighbour scheme} (Ahmad \& Cohen 1973) is
%key to a significant speed-up by performing the expensive
%calculation of the distant slowly changing force part (known as the
%``regular'' force) on GPUs. 
The code includes algorithms for stellar
and binary evolution (Tout et al.1997, Hurley et al.2000) and deals
directly with perturbations to binary orbits, collisions and
mergers, formation of three- and four-body subsystems, exchange
interactions and tidal capture. The treatment of close encounters
constitutes a large part of the code. 
%Two-body interactions are
%studied by the tools of Makino \& Aarseth (1992), preferentially by
%using a high-order Hermite scheme and an iterative solution without
%recalculating the external perturbation (Mikkola \& Aarseth 1993).
%In contrast, multiple encounters are treated by removing the
%two-body singularities via chain regularization (Mikkola \& Aarseth
%1993).
% The \texttt{NBODY6++} can be used in supercomputer thus it
%allow large $N$ simulation compared to $\nbodygpu$.

We set up a simple dynamical model of a massive stellar cluster, as
an example for {\tt GalevNB} to work on. The stellar system is
initially gas-free, and is assumed in virial equilibrium as an
approximation after losing gas (i.e. the ratio of
kinetic to potential energy is $Q_\mathrm{vir}$\,=\,0.5). 
The single-aged population is run with $N_0 = 10\,000$
single stars. Million-particle simulations of star clusters will be shown in our coming paper. 
The IMF is set up according to Kroupa (2001) with fixed lower and
upper mass limits, $m_\mathrm{l} = 0.1\,\Msun$ and $m_\mathrm{u} =
20\,\Msun$, respectively, resulting in an initial cluster mass of
$M_0 = 4.7\times 10^3\,\Msun$. 
To simplify the computation, stars are distributed according to
Plummer model with a scale radius of 0.59\,pc, which 
does not differ significantly from King model with $W_0=6$, except in
the outermost regions with low density.
The models have an initial half-mass
radius $\Rhm = 0.76$\,pc, and 
 is not primordially mass segregated.
We set the metallicity to sub-solar metallicity $Z_{\sun} = 0.001$,
which is similar to the halo population of Galactic globular
clusters. Note that our simulated
star cluster is still less massive than a typical Galactic globular
cluster. But here we are interested in special spectral features,
such as the production of UV bright stars, and aim to relate the
current simulations to future million particle simulations.

Since our primary interest is the
short-term stellar evolution, we therefore carry out a simulation of
3000 $N$-body time units (corresponding to 655.6\,Myr) with
dynamical and internal stellar evolution. Snapshots of the
simulation are selected to display in this paper for the presence of
UV-bright stars (see Section 4).

%Unlike Olczak et al.\ (2012), we neglect here the disruptive effect
%of a galactic potential as we are primarily interested in the
%long-term stellar evolution.

%Rotation has so far been neglected in simulations of young clusters
%though up to 30\,\% of the kinetic energy might be stored in a
%stable rotating mode (Einsel \& Spurzem 1999, Kim et al.\ 2002), as
%observed in some globular clusters see the review by Meylan \&
%Heggie (1997). We set up our model with $\omega_0 = 1.5$, the
%maximum for stable rotation (Einsel \& Spurzem 1999).

%Following Ernst et al. (2007), we add the effect of rotation via
%generalized two-parametric King models with King parameter $W_0 = -(
%\Phi_0 - \Phi_\mathrm{t}) / \sigma_\mathrm{K}^2$ and rotation
%parameter $\omega_0 = \sqrt{9/(4 \pi G \rho_\mathrm{c})} \Omega_0$,
%
%\begin{equation}
%  f( E, J_z ) = C \left[ \exp{\left( -\frac{E-\Phi_\mathrm{t}}{\sigma_\mathrm{K}^2} \right) - 1} \right] \exp{\left( -\frac{\Omega_0
%        J_z}{\sigma_\mathrm{K}^2} \right)} \,,
%\end{equation}
%
%where $E$ is the energy, $J_z$ the $z$-component of the angular
%momentum vector, $\Phi_0$ and $\Phi_\mathrm{t}$ the potential at the
%center and at the outer boundary (i.e. where the density approaches
%zero), $\sigma_\mathrm{K}$ the King velocity dispersion, $\Omega_0$
%is close to the angular speed in the cluster core, $G$ the
%gravitational constant, and $\rho_\mathrm{c}$ the central density.

\section{GalevNB: Galev for $N$-body simulations}
\label{sec:galevnb_sim}

\subsection{GalevNB Structure}
 In this section, we introduce
the \texttt{GalevNB} structure and its parameters. The main program
of \texttt{GalevNB} is \texttt{GalevNB.f90}, which parses single
snapshot files (stellar evolution only) generated by
\texttt{NBODY6++GPU} (\texttt{NBODY6++})/ \texttt{NBODY6GPU} (\texttt{NBODY6}). It invokes seven subroutines
(\texttt{startomaginit, specint\_initialize, reset\_weights,
 startomag, add\_star, spec2mag, spec\_output}) of \texttt{GALEV} package to convert effective
temperature, stellar luminosity, metallicity, and mass into
observational magnitudes and spectra. The functions of these
routines are presented in Table 1. The \texttt{GalevNB} package
contains four folders: 1) \texttt{spectral\_templates}, in which all
the spectral template files from the BaSeL library of model
atmospheres (Lejeune, Cuisinier \& Buser 1997, 1998) are stored; 2)
\texttt{standard\_filters}, containing a large set of filter
response functions (FUV, NUV, U, B, V, R, I, J, H, K) that are used
as standard reference filters; 3) \texttt{filter\_response\_curves},
including filter response functions from magnitude systems of HST,
ESO instruments, 2MASS, SDSS, Johnson, and Cousins in separate
subfolders. We also provide a choice of user-specify filter response
functions. Information about the entire set of available filters is
presented in the file \texttt{filterlist.dat}. Please aware that
\texttt{filterlist.dat}, in which the user specify their own choice
of magnitude system by uncommenting the row of chosen filter, MUST
be presented in the same directory as the \texttt{NBODY6++GPU} (\texttt{NBODY6++})/ \texttt{NBODY6GPU} (\texttt{NBODY6}) snapshot files. The content of the file,
\texttt{filterlist.dat}, is presented in Table 2.

\begin{table}[h]
\begin{center}
\caption[]{Functions of subroutines converting theoretical
parameters into observational magnitudes and spectra}\label{Table 1}
 \begin{tabular}{cc}
  \hline\noalign{\smallskip}
Subroutine  &  Function                 \\
  \hline\noalign{\smallskip}
\texttt{specint\_initialize}  &  initialize the stellar spectra       \\
\texttt{reset\_weights}   & reset the weight of stellar spectra             \\
\texttt{add\_star}  & integrate the flux of all stars in the cluster\\
\texttt{spec\_output} & output spectra \\
\texttt{startomaginit}  &  initialize the stellar magnitude\\ % new variable
\texttt{spec2mag}   &  convolve the stellar spectra with the filter response function\\
\texttt{startomag} &  compute magnitudes for stars\\
  \noalign{\smallskip}\hline
\end{tabular}
\end{center}
\end{table}

\begin{table}[h]
\begin{center}
\caption[]{Column contents for the filter information file:
\texttt{filterlist.dat}}\label{Table 1}
 \begin{tabular}{ccc}
  \hline\noalign{\smallskip}
Column  &  Content    & ID of zero point                  \\
  \hline\noalign{\smallskip}
1  & Filter name  &  \\ % new variable
2  & Corresponding path of the filter response function    &             \\
3  & ID of selected zero point (default value is 1)     &            \\
4  & Standard zero point in the Vega magnitude system & 1 \\
5  & Standard zero point in the AB magnitude system   & 2 \\
6  & Standard zero point in the ST magnitude system   & 3 \\
7  & Optional user-defined zero point                 & 4 \\
  \noalign{\smallskip}\hline
\end{tabular}
\end{center}
\end{table}

\subsection{Installation and Usage of GalevNB}
To compile \texttt{GalevNB}, the user should have {\tt C++} and
fortran installed. The input file of \texttt{GalevNB} should be a
sinlge snapshot output from \texttt{NBODY6++GPU} (\texttt{NBODY6++})/
\texttt{NBODY6GPU} (\texttt{NBODY6}) simulations. In case of a file containing all snapshots (called
\texttt{sev.83} in \texttt{NBODY6++GPU} / \texttt{NBODY6++} and \texttt{fort.83} in
\texttt{NBODY6GPU} / \texttt{NBODY6}), we provide the user with a shell script
\texttt{generate\_snapshots.sh} in the folder, \texttt{scripts}, for
retrieving single snapshot data out of
 \texttt{sev.83} and \texttt{fort.83}. The user can select his/her preferred filters
(maximum 20) by uncommenting the row of the corresponding filter in
\texttt{filterlist.dat}, and choose his/her desired magnitude system
(Table 2). Magnitudes of individual stars and the whole cluster, and
spectra of the cluster or chosen stellar types are produced,
respectively. The code of \texttt{GalevNB} is published online\footnote{\tt http://silkroad.bao.ac.cn/repos/galevnb}.
Users can download it through internet. We also provide examples
of output files from \texttt{NBODY6++GPU}  and \texttt{GalevNB} on the
web\footnote{\tt http://silkroad.bao.ac.cn/\textasciitilde xiaoying/}.
% \footnotetext{\tt \url{http://silkroad.bao.ac.cn/~xiaoying/}} 

\section{Observational features of simulated star clusters}
\label{sec:obs}

 With \texttt{NBODY6++GPU} simulations done in Section 2, and
the computation of \texttt{GalevNB}, we are able to investigate
early evolution of CMD and SED of a stellar population in a real
dynamical environment.
 The output of \texttt{NBODY6++GPU}
simulations contain integers to represent stellar type information
(Hurley et al. 2000) of each individual stars (Table 3). This allows
one to identify exotic objects in the star cluster based on the
evolution of CMD and SED. It is true that cluster dynamics, such as
close stellar encounters or the formation of binaries, can
significantly alter the stellar populations in dense stellar systems
(Djorgovski \& Piotto 1993). Nevertheless, to simplify the
\texttt{NBODY6++GPU} simulations that \texttt{GalevNB} works on, we do
not consider primordial binaries in the current simulations.
Therefore, stellar type 19-22 (binaries) do not present in our  
simulations.
 Due to the faintness of stellar remnant, stellar types 13-15 (though appear
 in our examples) are not included in the studies of CMD and SED.

\begin{table}[h]
\begin{center}
\caption[]{Stellar type defined in the \texttt{NBODY6++GPU}
codes}\label{Table 3}
 \begin{tabular}{cc}
  \hline\noalign{\smallskip}
 Integer
  &  Stellar type                \\
  \hline\noalign{\smallskip}
0  & main sequence (MS) stars with mass$\le0.7\,M\odot$  \\ % new variable
1  & MS with mass$\ge0.7\,M\odot$            \\
2  & Hertzsprung Gap          \\
3  & Red Giant \\
4  & Core Helium Burning \\
5  & First (early) Asymptotic Giant Branch \\
6  & Second Asymptotic Giant Branch ($\sim$Thermal Pulsing AGB) \\
7  & Naked Helium Star MS \\
8  & Naked Helium Star Hertzsprung Gap \\
9  & Naked Helium Star Giant Branch  \\
10 & Helium white dwarf  \\
11 & Carbon/Oxygen White Dwarf  \\
12 & Oxygen/Neon White Dwarf  \\
13 & Neutron Star  \\
14 & Black Hole \\
15 & Massless supernova remnant \\
19 & Circularizing binary (c.m. value) \\
20 & Circularized binary \\
21 & First Roche stage (inactive) \\
22 & Second Roche stage \\
  \noalign{\smallskip}\hline
\end{tabular}
\end{center}
\end{table}
%Description of Figure 1.

\subsection{SED \& CMD evolution}

After integrating fluxes of individual stars, \texttt{GalevNB}
output total SED of the star cluster. In the main program, {\tt
GalevNB.f90}, user can choose to output SED of certain stellar
types. Therefore, specific stars' contribution to the total SED
of the cluster can be quantified. In Figures 1, 3, 5, 7, we display
SEDs of the cluster (black line), of asymptotic giant branch
(AGB) stars (red line), of naked Helium stars (green line), of white
dwarfs (blue), and of core helium burning stars (cyan line) at different
ages. Movies of the evolution of SED and CMD are available in the weblink provided in Section 3.2$^2$. 

At very early age, massive stars first exhaust their hydrogen in the
core and ignite core helium burning. These core helium burning stars
are UV-bright (Figure 1) and very blue (Figure 2), which are also
termed as blue loops (Stothers \& Chin 1991, El Eid 1995). The UV
peak in the SED of the cluster ($<1000$\,\AA) is mainly due to the
presence of these stars (Figure 1).

As the cluster evolves, the intensity of the UV-excess drops (Figure
3, 5, 7). Besides core helium burning stars, we find another three
stellar types producing UV radiation, second AGB stars, white dwarfs,
and naked helium stars.

1) Early AGB stars (stellar type = 5) are very red, and therefore
mainly radiate in the red filters (Girardi 2000, Marigo et al. 2008,
Salaris et al. 2014). After the third dredge-up, helium shells flash
repeats many times. At this stage, early AGB becomes thermal pulsing
AGB (TPAGB; Marigo et al. 2008, Hurley et al. 2000). In the 
\texttt{NBODY6++GPU} code, the TPAGB are called "second AGB" (stellar
type = 6). The radii of the TPAGB grow so large that mass-loss is
significant.
 At this time, the TPAGB reach very high temperature ($>10^4$\,K) and begin to be
  luminous in the UV filters (Figure 3). Mass-loss will eventually
remove all the envelopes of the TPAGB, which may be seen as
planetary nebulae (see the CMD in Figure 4).

2) White dwarfs are born after the TPAGB turning into planetary
nebulae. Though white dwarfs cool down eventually
 (Mestel 1952, Liebert 1980, Hansen \& Liebert 2003), young white dwarfs are very hot and blue (Mestel
 1952, Rauch et al. 2014, Torres et al. 2014), whose radiation is
largely at the UV ($<2000$\,\AA, Figures 5 \& 7). Even though their
luminosity at the UV is not high, their contribution to the
UV-excess of the cluster is long-term and might not be neglected,
since stars evolve to white dwarfs continually.

3) One short term radiator at the UV is the naked helium star, with
extremely blue color over the main-sequence turn-off (Figures 7 \&
8). They are massive stars losing hydrogen envelopes, the so-called
Wolf-Rayet stars in observations (Crowther 2007, Shara et al. 2013).
Since naked helium stars only appear at young age for a short
time-scale, they cannot influence the UV-excess of the cluster in
the long run.

\begin{figure}[t]
\centering
\includegraphics[width=13cm]{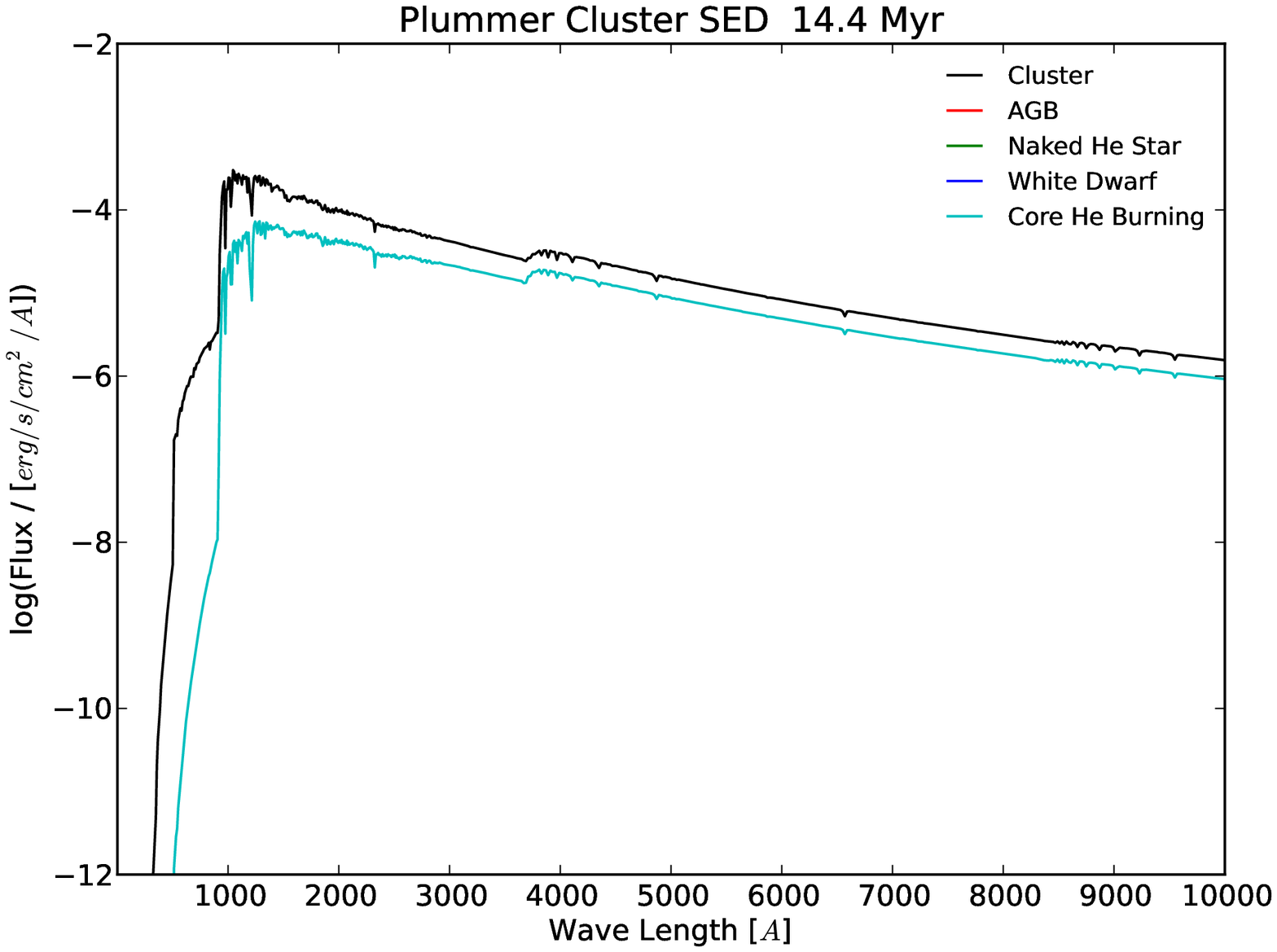}
\caption{SED of the simulated cluster (black line) and core helium
  buring stars (cyan line) at the age of
14.4\,Myr. SED of other stellar types will be shown in
the following figures. SED of core helium burning stars (stellar
type = 4) is indicated as cyan line, AGB stars (stellar type = 5 \&
6) red line, naked helium stars (stellar type = 7, 8, 9) green line,
white dwarfs (stellar type = 10, 11, 12) blue line. High resolution
figures are available for online version. \label{Fig.1}}
\includegraphics[width=13cm]{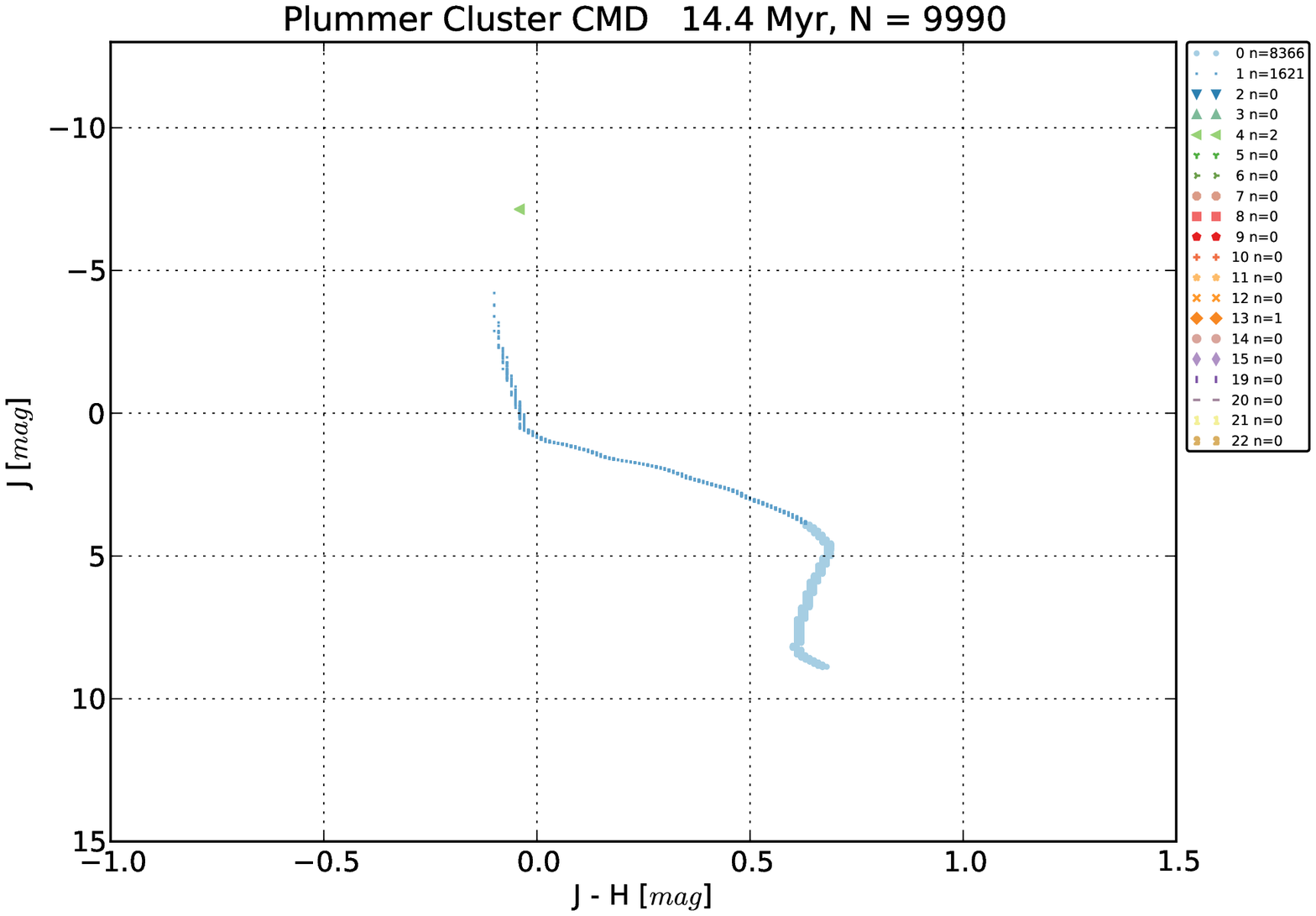}
\caption{J-H Color-Magnitude diagram of the simulated cluster at the
age of 14.4\,Myr. N is the total number of stars, and n is the
number of stars of each stellar type indicated in the separate box to
the right. Stellar types (see Table 3) are
marked with different color and by different symbols. 
Main sequence stars (Light blue filled round dot: stellar type = 0; sky blue one-pixel point: stellar
type =1) and core helium burning stars (light green
left-pointing-triangle: stellar type = 4) 
appear in this CMD.  High resolution
figures are available for online version.\label{Fig.2}}
\end{figure}

\begin{figure}[b]
\centering
\includegraphics[width=13cm]{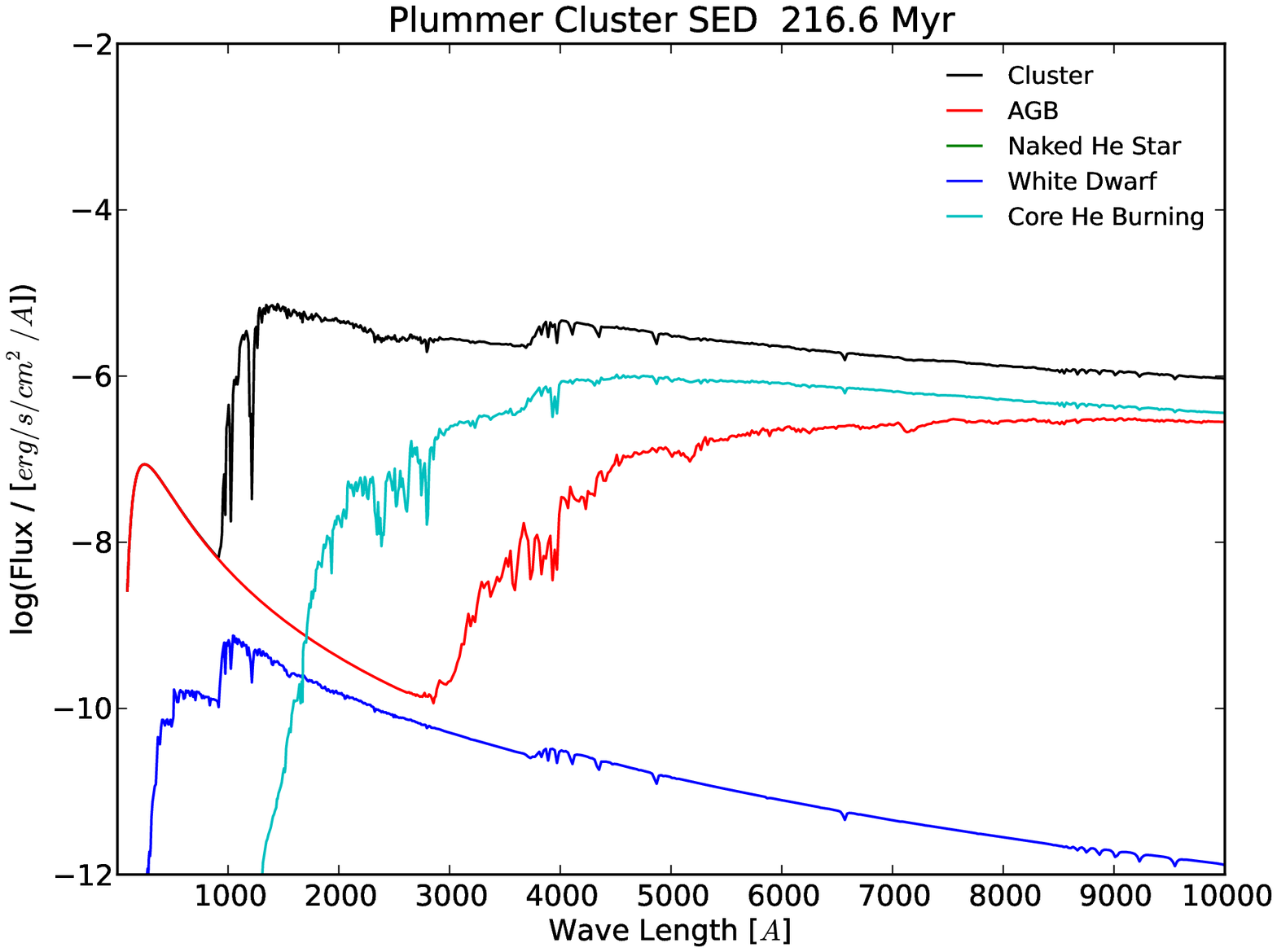}
\caption{SED of the simulated cluster (black line), AGB
  stars (red line), core helium burning stars (cyan line) and white
  dwarfs (blue line) at the age of
216.6\,Myr. Color coding is the same as Figure 1. High resolution
figures are available for online version.
\label{Fig.3}}
\includegraphics[width=13cm]{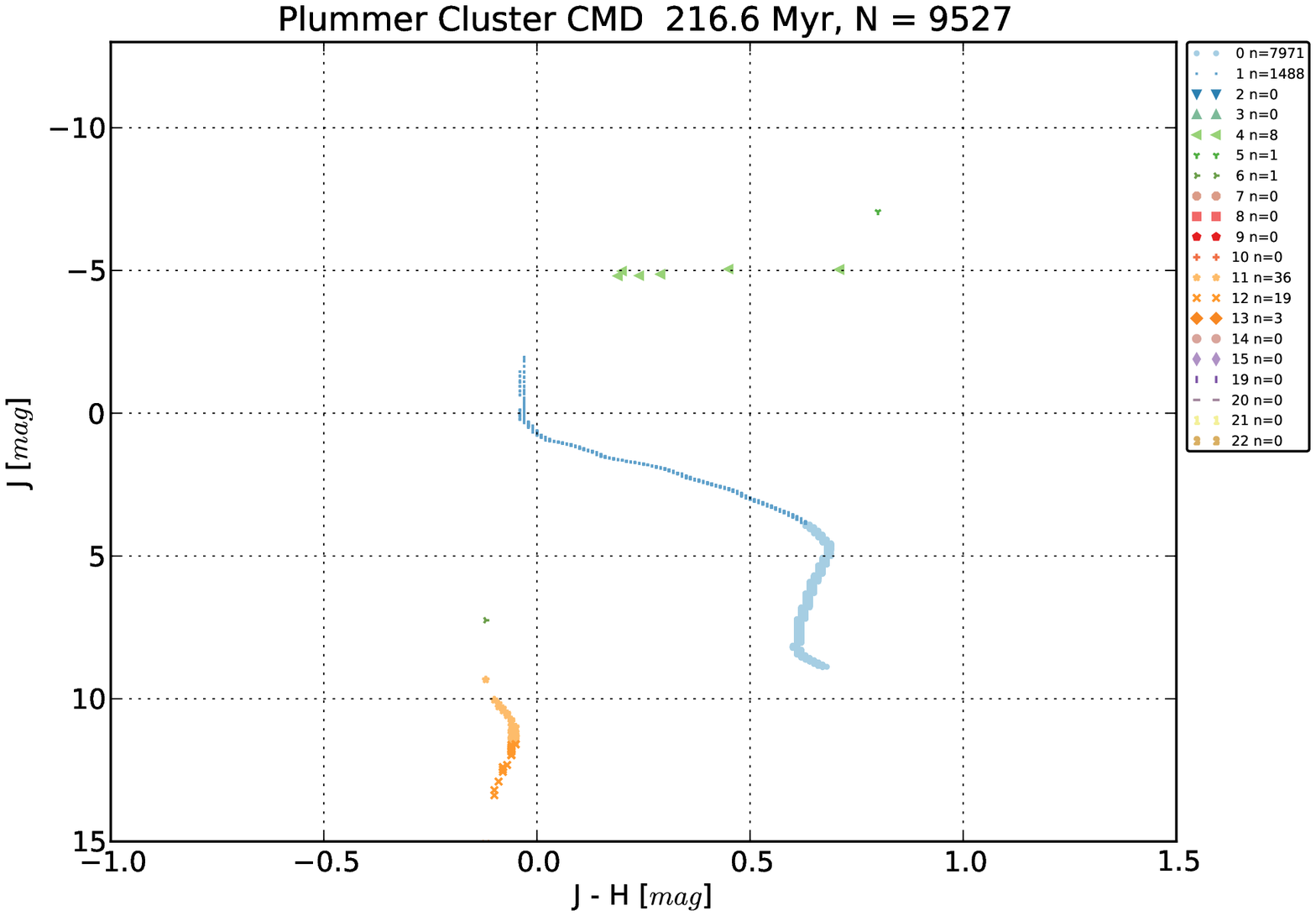}
\caption{J-H Color-Magnitude diagram of the simulated cluster at the
age of 216.6\,Myr.  N is the total number of stars, and n is the
number of stars of each stellar type indicated in the separate box to
the right. Stellar types (see Table 3) are
marked with different color and by different symbols. Main sequence stars (Light blue filled round dot: stellar type = 0; sky blue one-pixel point: stellar
type =1), core helium burning stars (light green
left-pointing-triangle: stellar type = 4), AGB stars (green
downward-tripple-point: stellar type =5; green upward-tripple-point:
stellar type =6) and white dwarfs (yellow star: stellar type =11;
orange x-cross: stellar type =12)
appear in this CMD. High resolution
figures are available for online version.\label{Fig.4}}
\end{figure}

\begin{figure}[t]
\centering
\includegraphics[width=13cm]{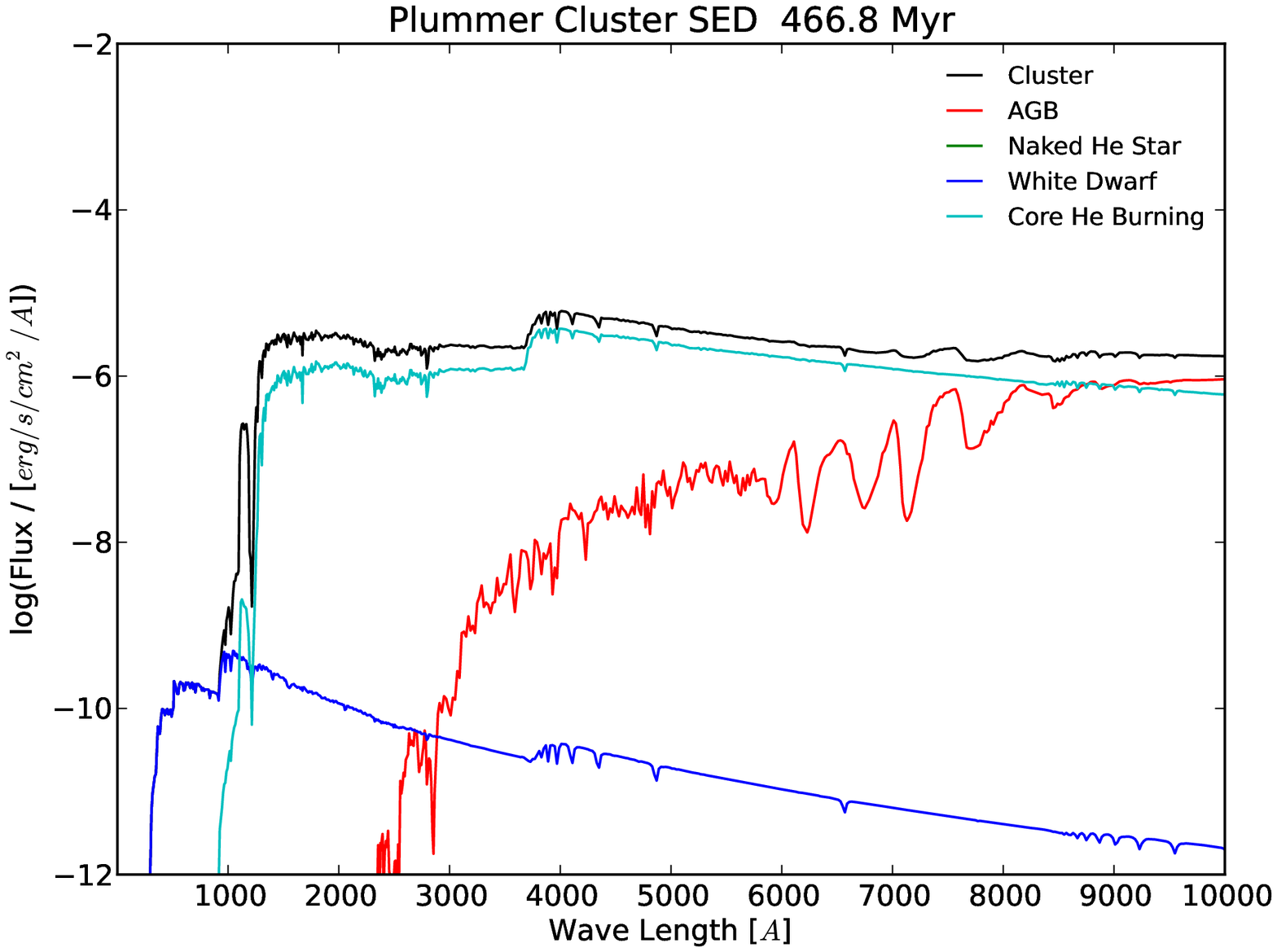}
\caption{SED of the simulated cluster (black line), AGB
  stars (red line), core helium burning stars (cyan line) and white
  dwarfs (blue line) at the age of
466.8\,Myr. Color coding is the same as Figure 1. High resolution
figures are available for online version.
\label{Fig.5}}
\includegraphics[width=13cm]{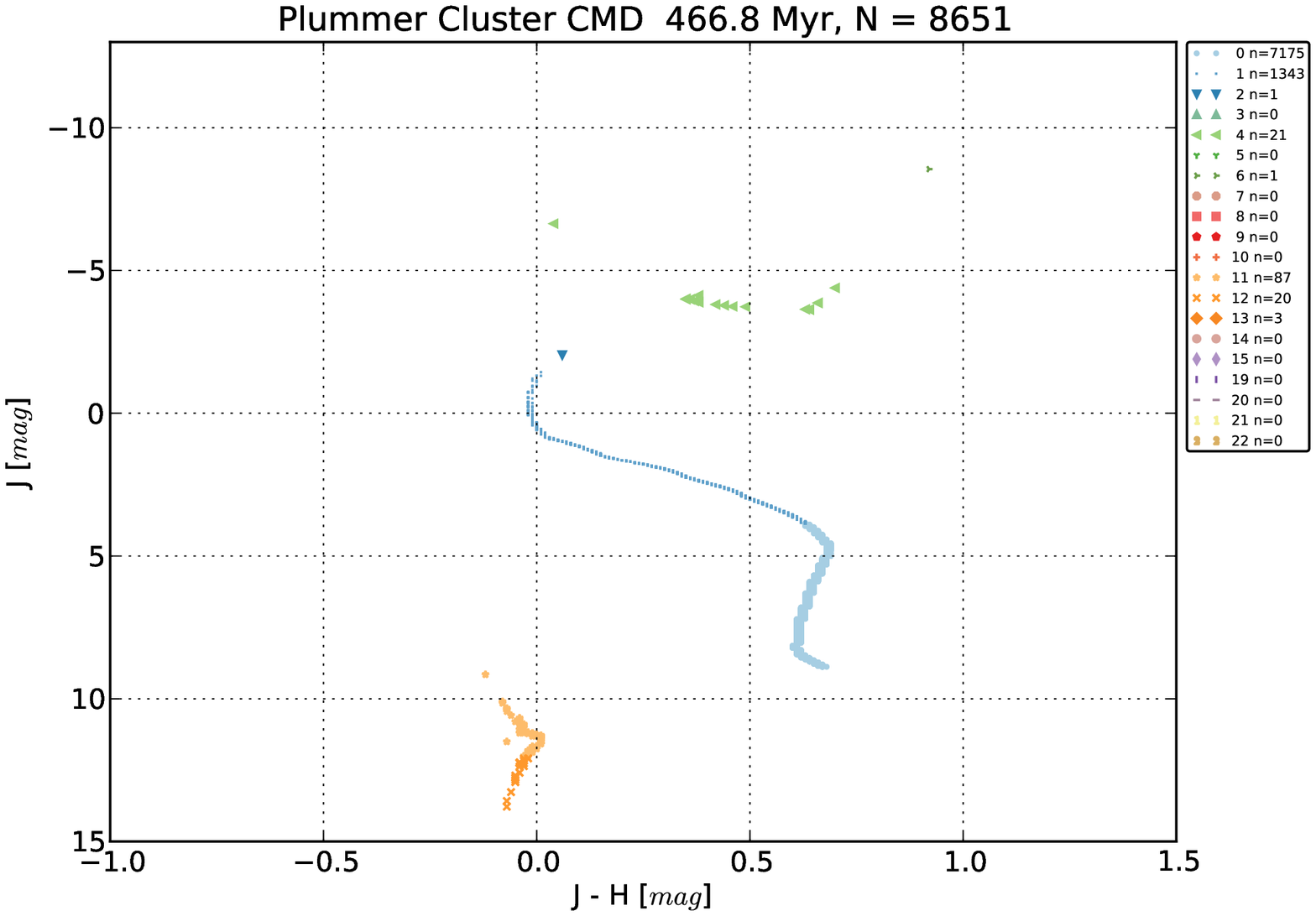}
\caption{J-H Color-Magnitude diagram of the simulated cluster at the
age of 466.8\,Myr.  N is the total number of stars, and n is the
number of stars of each stellar type indicated in the separate box to
the right. Stellar types (see Table 3) are
marked with different color and by different symbols. Main sequence stars (Light blue filled round dot: stellar type = 0; sky blue one-pixel point: stellar
type =1), Hertzsprung gap stars (dark blue filled downward-triangle:
stellar type =2), core helium burning stars (light green
left-pointing-triangle: stellar type = 4), AGB stars (green upward-tripple-point:
stellar type =6) and white dwarfs (yellow star: stellar type =11;
orange x-cross: stellar type =12)
appear in this CMD. High resolution
figures are available for online version. \label{Fig.6}}
\end{figure}

\begin{figure}[b]
\centering
\includegraphics[width=13cm]{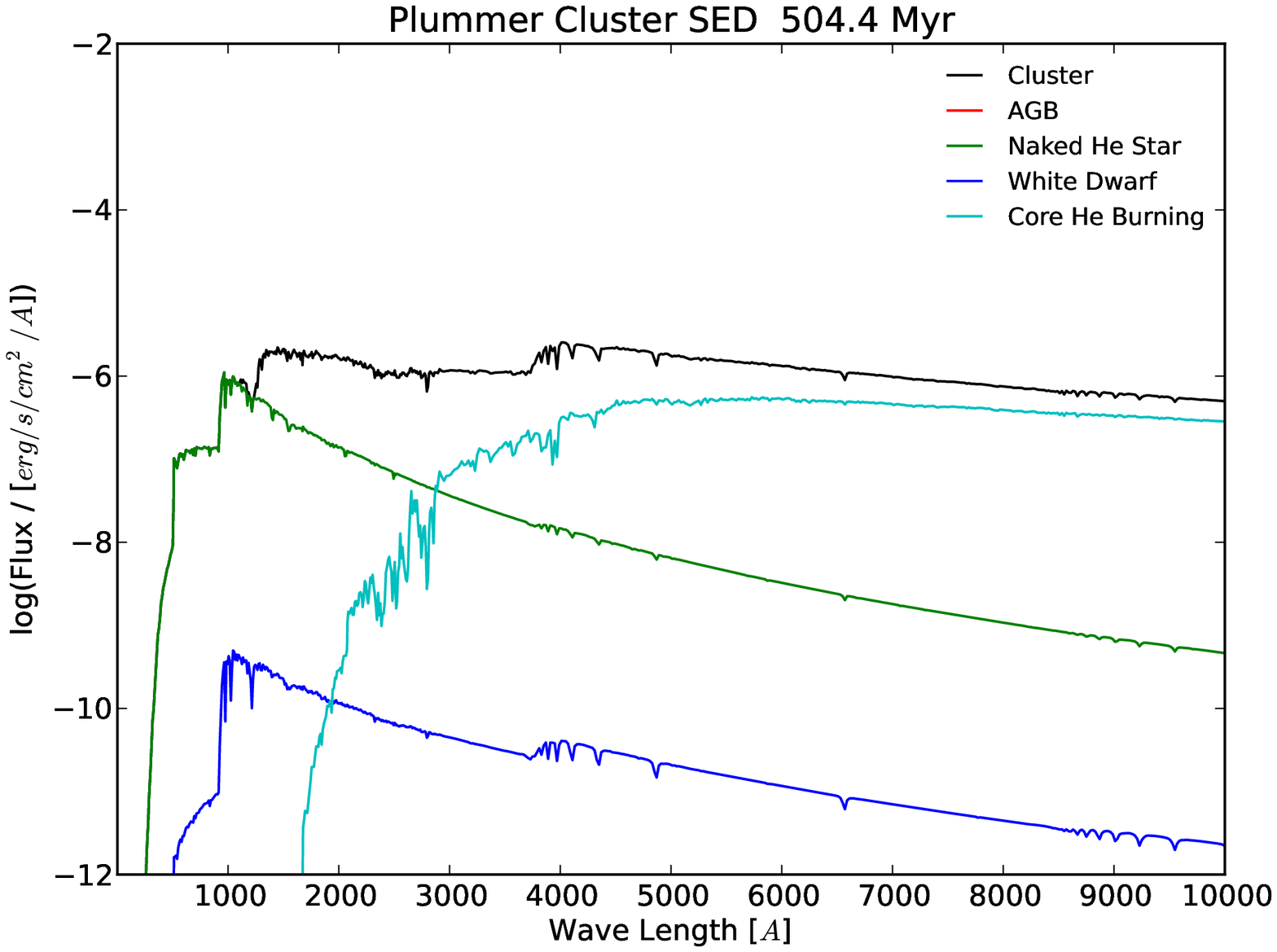}
\caption{SED of the simulated cluster (black line)
  naked helium stars (green line), core helium burning stars (cyan line) and white
  dwarfs (blue line) at the age of
504.4\,Myr. Color coding is the same as Figure 1. High resolution
figures are available for online version.
\label{Fig.7}}
\includegraphics[width=13cm]{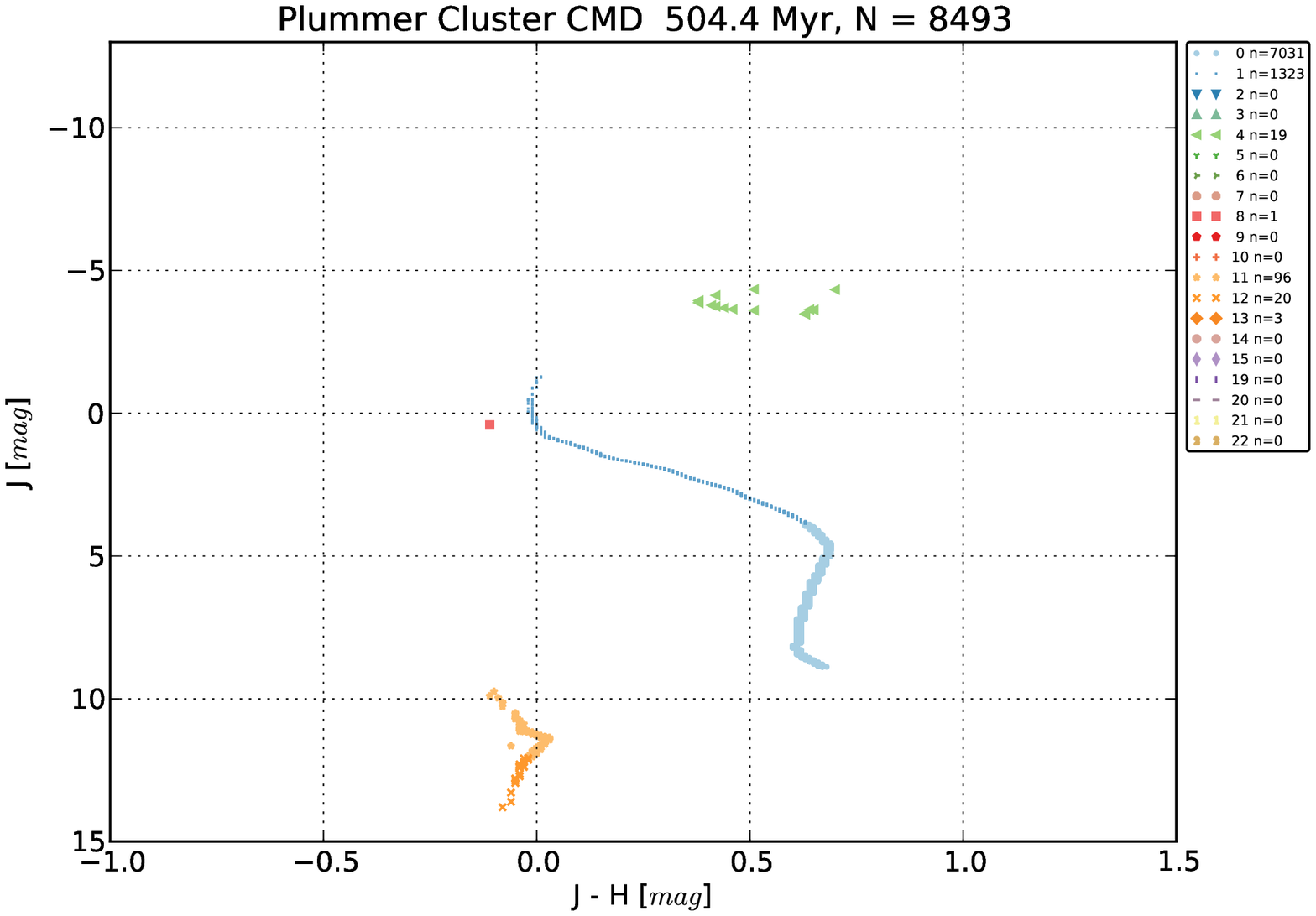}
\caption{J-H Color-Magnitude diagram of the simulated cluster at the
age of 504.4\,Myr.  N is the total number of stars, and n is the
number of stars of each stellar type indicated in the separate box to
the right. Stellar types (see Table 3) are
marked with different color and by different symbols. Main sequence stars (Light blue filled round dot: stellar type = 0; sky blue one-pixel point: stellar
type =1), naked helium burning stars (pink-red square: stellar type =
8), core helium burning stars (light green left-pointing-triangle: stellar type = 4), and white dwarfs (yellow star: stellar type =11;
orange x-cross: stellar type =12)
appear in this CMD. High resolution
figures are available for online version. \label{Fig.8}}
\end{figure}

\section{Discussion}
\label{sec:dis}

We present an integrated software solution, \texttt{GalevNB}, a
translator that bridges \texttt{NBODY (6++ / 6)} simulations (with or
without GPU) and
observations. We run a short-term dynamical model of a star cluster
based on \texttt{NBODY6++GPU} code, producing data for
\texttt{GalevNB}. \texttt{GalevNB} computes observational magnitudes
and spectra of stars by theoretical parameters, stellar mass,
temperature, luminosity and metallicity, which are generated by
\texttt{NBODY6++GPU} simulations. In the SED evolution of the simulated
cluster, we found a UV-excess in the wavelength shorter than
2000\,\AA.

Similar phenomenon is found in elliptical galaxies and early-type
spiral bulges whose SEDs surprisingly increase to shorter
wavelengths over the range 2000 to 1200 \AA, (Code 1969, Code et
al.\ 1972, Faber 1983, Burstein et al.\ 1988, Kurucz 1991), in
contrary to the expectation of an old stellar systems that are
usually assumed to be dark in the FUV. This rise in the spectrum at
wavelengths shorter than 2000\,$\rm \AA$ is called "UV upturn'' or
"UV-excess''. The UV-excess resembles the Rayleigh-Jeans tail of a
thermal source with effective temperature larger than 20000\,K
(Hills 1971). During the last decades, many efforts have been made
to find out "culprits'' of the UV-excess. In contrast to our
simulated cluster whose age is young (up to 655.6\,Myr), in most of
the early-type galaxies young massive stars are absent (O'connell
1986, Welch 1982, Buzzoni et al.\ 2012). Therefore, old, hot,
low-mass stars become a more popular choice.
 Extreme blue horizontal branch stars, with core helium
burning and very thin envelopes, might be promising for explaining
this phenomenon. These stars are found in both metal-rich and
metal-poor star clusters (Rich et al. 1997, Ferarro 1998, Li et al.
2013, Buzzoni et al. 2012), which makes their origins somehow
ambiguous. Theorists proposed another candidate producing a
significant amount of UV radiation, post-AGB stars (Buzzoni \&
Gonzalez-Lopezlira 2008), a quarter of which undergo TPAGB phase
(Lawlor \& MacDonald 2006). Some studies suggested that new-born
white dwarfs were hot enough to emit moderate amount of UV photons
(Hills 1971, Bica et al.\ 1996b). Recently, based on detail
computations of binaries, Han et al.\ (2007) conclude that the brown
dwarf binary should be a more promising candidate accounting for the
UV-excess.

Most of the candidate stars mentioned above originate in the cluster
environment. With the application of \texttt{GalevNB} to
\texttt{NBODY6++GPU} simulation data, we are able to observe the SED
evolution of different stellar types. Through this way, we find out four
kinds of hot, blue stars that are dominant in the UV-excess: core
helium burning stars, second AGB (TPAGB), white dwarfs and naked
helium stars. Among them, second AGB is a favorable candidate from
theoretical point of view (O'Connell 1999). On the contrary, white
dwarf's candidate position is controversial (Magris \& Bruzual 1993,
Landsman et al.\ 1998) because of low luminosity. The life time of
naked helium stars is very short. Though they are very bright at the
UV, their short-term emission makes them insignificant candidates.
However, how the UV radiation of the candidate stars evolve in the
long-term is beyond the scope of this paper, which cannot be
achieved with a simple stellar model. A detailed \texttt{NBODY6++GPU}
simulation investigating the evolution of UV-excess in star
clusters, with more particles, realistic initial condition and
primordial binaries, will be presented in our coming paper (Pang et
al. in prep.). The \texttt{NBODY6++GPU} code in its latest GPU
accelerated version is publicly available in Wang et al.
2015. We provide our scripts and subroutines of \texttt{GalevNB} online$^1$.  Please notice
that these are compatible with both \texttt{NBODY6++ \& 6} (with or
without GPU)
version. Users of the other version of \texttt{NBODY} may have to
adjust some data formats.

   For more than a decade ''simulating observations of simulations'' has been a
topic in the community of MODEST (MODelling Dense Star Clusters, see
\texttt{www.manybody.org/modest/} and Kouwenhoven et al. 2004).
Hurley et al. (2005) have published a pioneering study, which
presented a full Hertzsprung-Russell diagram (luminosity vs.
effective temperature) of all stars in the cluster M67 obtained from
a direct $N$-body simulation with \texttt{NBODY6}, to be directly
comparable to observations. These capabilities come with the
standard public versions of \texttt{NBODY6} and \texttt{NBODY6++}
(with or without GPU) already. The Monte Carlo code \texttt{MOCCA} is now also providing
similar capabilities (Giersz, Heggie \& Hurley 2008);
  recently it has been extended to allow simulated observations of star clusters
with specific telescopes using their colour windows and standard
software used by observers (Askar et al. 2015).
  This work has been driven by future key observational projects (for example by
Hubble Space Telescope) for globular clusters (e.g. Milone et al.
2014); however, to fully uncover the 6D dynamical structure and the
  3D gravitational potential, as well as the population history of star
  clusters, pure photometry may not be sufficient. Future deep high resolutions
spectroscopic and spectrophotometric observations of globular and
other star clusters are required in order to uncover one more
dimension in velocity (radial velocity) and obtain more information
about the stellar population through full spectra. Kacharov et al.
(2014) and den Brok et al. (2014) provide examples on how kinematic
data provide key insights into the dynamics of globular
  clusters (rotation and the quest for intermediate mass black holes). See also
for an overview van de Ven (2010). Here the combination of the
\texttt{GALEV} codes, previously mainly used for synthesis
  of galactic stellar populations (Kotulla et al. 2009) with our most recent
direct $N$-body code for star clusters is a pioneering step to
provide the corresponding full spectral information at any time and
any
  region in the cluster, star-by-star or in integrated fields from the models.

\begin{acknowledgements}
This work is funded by National Natural Science Foundation of China,
No: 11443001 (XYP) and 11073025 (RS). XYP extends gratitute to the
funds of National Natural Science Foundation of China, No: 11503015,
and of Shanghai education committee, No: 1021ZK151009027-ZZyy15104,
and of the talents introduction project of Shanghai Institute of Technology, No:
10120K156031-YJ2014-05. 
XYP and RS are members of the
Silk Road Project Team in National Astronomical Observatories of
China (NAOC, \texttt{http://silkroad.bao.ac.cn}), and acknowledge
the technical support from this team. CO appreciates funding by the
German Research Foundation (DFG) grant OL 350/1-1; CO and RS have
been partly supported through computational resources of SFB 881
¡°The Milky Way System¡± (subproject Z2) at the University of
Heidelberg, Germany, in particular the Milky Way supercomputer
hosted and co-funded by the J\"ulich Supercomputing Center (JSC).
XYP appreciates the travel grants of the DFG grant OL 350/1-1. 

RS is grateful to support by the Chinese Academy of Sciences
Visiting Professorship for Senior International Scientists, Grant
Number 2009S1-5, and through the "Qianren'' special foreign experts
program of China, both at NAOC.
  The special GPU accelerated supercomputer \texttt{laohu} at the Center of Information and
Computing at National Astronomical Observatories, Chinese Academy of
Sciences, funded by Ministry of Finance of People¡¯s Republic of
China under the grant ZDY Z2008-2, has been used for simulations, as
well as smaller GPU clusters titan, hydra and kepler, funded under
the grants I/80041-043 and I/84678/84680 of the Volkswagen
Foundation at ARI/ZAH, University of Heidelberg, Germany.

R.K. gratefully acknowledges Financial support from the National
Science Foundation under Grant No. 1412449, as well as from STScI
theory grant HST-AR-12840.01-A.

%We are grateful to Dr. Ralf Kotulla for kindly offering the original
%\texttt{GALEV} subroutines for this project, for his early
%initiative on this project and the subsequent useful discussion. 
We appreciate Dr. Peter Anders for promoting and coordinating the
collaborations between our $N$-body team and the \texttt{GALEV}
team.
Our gratitude also to Long Wang and Maxwell Xu Cai for discussion
and support in running simulations and programming. We also
acknowledge Prof. Dr. Richard de Grijs from Peking University and
Prof. Dr. Zhongmu Li from Dali University for helpful discussion
during this work.

%   Milky Way
%  supercomputer, which is funded by the Deutsche Forschungsgemeinschaft (DFG) through the Collaborative Research Center (SFB 881) ``The Milky Way
%  System'' (subproject Z2) and hosted and co-funded by the J\"ulich Supercomputing Center
%  (JSC).
\end{acknowledgements}

%\appendix                  %%appendicial material is supported

\label{lastpage}

\end{document}